\documentclass[Afour,sagev,times,fleqn]{sagej}

\usepackage{moreverb,url}

\usepackage[colorlinks,bookmarksopen,bookmarksnumbered,citecolor=red,urlcolor=red]{hyperref}

\usepackage{algorithm, algcompatible, setspace, amsmath}
\usepackage{float}
\usepackage{flushend}

\newcommand{\RNum}[1]{\uppercase\expandafter{\romannumeral #1\relax}}

\newcommand{\zcedit}[1]{{\color{black} #1}}

\newcommand\BibTeX{{\rmfamily B\kern-.05em \textsc{i\kern-.025em b}\kern-.08em
T\kern-.1667em\lower.7ex\hbox{E}\kern-.125emX}}

\def\volumeyear{2020}

\begin{document}

\runninghead{Chen and Sun}

\title{Sparse representation for damage identification of structural systems}

\author{Zhao Chen\affilnum{1} and Hao Sun\affilnum{1, 2}}

\affiliation{\affilnum{1}Department of Civil and Environmental Engineering, Northeastern University, Boston, MA, USA\\
\affilnum{2}Department of Civil and Environmental Engineering, Massachusetts Institute of Technology, Cambridge, MA, USA}

\corrauth{Hao Sun, Department of Civil and Environmental Engineering, Northeastern University, Boston, MA 02115, USA.}

\email{h.sun@northeastern.edu}

\begin{abstract}
Identifying damage of structural systems is typically characterized as an inverse problem which might be ill-conditioned due to aleatory and epistemic uncertainties induced by measurement noise and modeling error. Sparse representation can be used to perform inverse analysis \zcedit{for the case of sparse damage.} In this paper, we propose a novel two-stage sensitivity analysis-based framework for both model updating and sparse damage identification. Specifically, an $\ell_2$ Bayesian learning method is firstly developed for updating the intact model and uncertainty quantification so as to set forward a baseline for damage detection. A sparse representation pipeline built on a quasi-$\ell_0$ method, e.g., Sequential Threshold Least Squares (STLS) regression, is then presented for damage localization and quantification. Additionally, Bayesian optimization together with cross validation is developed to \zcedit{heuristically learn hyperparameters from data}, which saves the computational cost of hyperparameter tuning and produces more reliable identification result. The proposed framework is verified by three examples, including a 10-story shear-type building, a complex truss structure, and a shake table test of an eight-story steel frame. Results show that the proposed approach is capable of both localizing and quantifying structural damage with high accuracy.
\end{abstract}


\maketitle

\section{Introduction}
Model updating is an effective way to address the discrepancy between an ideal finite element model and the actual system based on sensing data. Such a discrepancy might be attributed to measurement noise and/or modeling error. Updated models can then \zcedit{be used to} predict structural response, \zcedit{identify} damage, and perform reliability analysis, among others. The core idea of model updating is to find the representative variation of structural properties that can account for the location and the extent of discrepancies \cite{Mottershead}. Among many classical methods (e.g. least squares-based method \cite{Xu}, heuristic algorithm \cite{Sun2013}, filtering techniques \cite{Chatzi}, etc.), sensitivity analysis is one of the most mature methods for modeling updating \cite{Mottershead, Marwala}. However, a distinctive challenge is that extracting structural parameters from measurement data, such as modal frequencies and shapes, is typically an ill-posed regression problem in the context of sensitivity analysis, due to (1) measurement incompleteness and noise, and (2) inevitable modeling error. 

To tackle this issue, regularization has been applied, among which the Tikhonov regularization \cite{Tikhonov} and the truncated singular value decomposition \cite{Hansen} are very popular. Besides the deterministic methods, Bayesian learning has been adopted to quantify uncertainties associated with model updating \cite{Beck}. Lots of theoretical and experimental contributions have \zcedit{advanced the methodology} for model updating and damage identification. To name a few: a two-stage modal-based Bayesian model updating strategy with ambient measurements \cite{Zhang}; a hierarchical Bayesian modeling that accounts for time-variability of structural systems \cite{Behmanesh}; Bayesian inference for simultaneous identification of structural parameters and loads \cite{Sun2015}; damage detection of shear frames with scarce and noisy measurements \cite{Sohn}; multiresolution Bayesian regression for model updating which can flexibly zoom \zcedit{into} significant regions \cite{Yuen2018}; recursive Bayesian updating \zcedit{that employs} frequency response function \cite{Mao}; bolted-connection damage detection using incomplete and noisy modal data \cite{Yin2017}; Bayesian damage prognosis for remaining useful life of bearings \cite{Mao2014}; identification of the Phase \zcedit{\RNum{2}} ASCE-IASC benchmark frame \cite{Ching}; and progressive damage identification of a 7-story building \zcedit{slice} \cite{Simoen}. Comprehensive literature reviews on this topic are well elaborated in \cite{Au,Yuen,Simoen2015, Huang2019}.

While there have been significant developments for solving the inverse problem \zcedit{for} model updating, most of existing approaches are based on $\ell_2$ regularization, which tend to ``over-smooth'' structural parameter variations. \zcedit{While these methods are useful for largely populated damages, for example, widespread surface corrosion, they tend to lead to biased damage identification in scenarios where damage has sparsity characteristics (e.g., damage occurs at distinct locations of the structure with $\ell_0$ features from a mathematical point of view)}. Therefore, sparsity-promoting regularization is required \zcedit{for such a situation}. Recently, Huang \emph{et al.} \cite{Huang2015,Huang,Huang2017b} developed a group of sparse damage identification methods based on Bayesian $\ell_1$ learning that imposes spatially sparse constraints on ill-proposed model updating problems with incomplete modal data. Nevertheless, the $\ell_1$-oriented sparse Bayesian learning is \zcedit{still} suffered from false positives in identification, since the $\ell_1$-norm is a relaxation of sparsity and serves as an approximation of $\ell_0$ regularization \zcedit{unless a certain strong condition is satisfied \cite{Candes}}. Despite of its ideal characteristics for sparse representation, the non-differentiability of the $\ell_0$ norm makes its optimization a non-deterministic polynomial-time (NP) hard problem, possessing extremely computational complexity and preventing its wide application. Recently, the Sequential Threshold Least Squares (STLS) regression proposed by Rudy \emph{et al.} \cite{Rudy} has shown efficient and superior sparsity representation in the context of $\ell_0$ regularization based on selective hard-thresholding. Though this approach shows good promise for solving sparse damage identification problems, a critical issue lies in how to choose the thresholding criterion. It is noted that this criterion is very problem-dependent and an inappropriate selection will essentially lead to biased identification. We will address this fundamental issue by introducing an automatic thresholding mechanism.

In this paper, we propose a two-stage sensitivity analysis-based framework for model updating and sparse damage identification, cohesively combining $\ell_2$ and $\ell_0$ regularization. In the model updating stage, an $\ell_2$ Bayesian learning method is firstly developed for model updating and quantifying associated model parameter uncertainties. The updated model will serve as a baseline for damage detection. In the sparse damage identification stage, a quasi-$\ell_0$ method based on STLS regression is utilized along with cross validation and Bayesian hyperparameter optimization to enable \zcedit{a data-driven} sparse representation of damage.

The rest organization of this paper is as follows. The second Section presents the methodology of the model updating ($\ell_2$ Bayesian learning) and the sparse damage identification (STLS regression). The third Section shows two numerical and one experimental examples to verify the performance of the proposed approach. The fourth Section summarizes the conclusion of this paper.

\section{Methodology}\label{Method}
This section presents the framework of sensitivity analysis-based model updating and sparse damage identification. The first subsection introduces the concept of sensitivity analysis which formulates the problem of nonlinear model updating in a recursive process. The next subsection introduces $\ell_2$ Bayesian learning for solving the resulting sensitivity equation. The final subsection elaborates STLS regression for damage detection and its enhancement by cross validation and Bayesian hyperparameter estimation.

\subsection{Sensitivity Analysis}
Sensitivity analysis is to tell the influence of latent factors on the observable behavior of a system. To begin with, we parameterize the structural model with respect to the stiffness parameters at the local element level, namely, 
\begin{equation} \label{eq:StiffnessParameterization}
\mathbf{K}_1 = \mathbf{K}_0 + \sum_{i=1}^{N_{ele}} \theta^i \mathbf{K}_0^i
\end{equation}
where $\mathbf{K}_0 \left( \in \mathbb{R} ^{N_{DOF} \times N_{DOF}} \right)$ is the stiffness matrix of the initial model, $\mathbf{K}_1 \left( \in \mathbb{R} ^{N_{DOF} \times N_{DOF}} \right)$ is the stiffness matrix of the updated model, $N_{ele}$ is the number of structural elements, $N_{DOF}$ is the total number of degrees-of-freedom (DOFs), and $\mathbf{K}_0^i \left( \in \mathbb{R} ^{N_{DOF} \times N_{DOF}} \right)$ is the stiffness matrix of $i$th substructure, which is derived from the secondary diagonal of $\mathbf{K}_0$. Lastly, $\theta_i \left( \in \mathbb{R} ^{N_{ele} \times 1} \right)$ is the variation coefficient of $\mathbf{K}_0^i$. Combining all $\theta_i$, we have a variable vector for all structural elements $\boldsymbol{\theta} = [ \theta^1, ..., \theta^i, ...\theta^{N_{ele}} ]^\top$.

The following sensitivity equation, which connects the modal residue (of frequencies and shapes) and the structural parameters, is derived as a basis for model updating:
\begin{equation} \label{eq:SensitivityEquation}
\mathbf{r}_k = \mathbf{S}_k \Delta \boldsymbol{\theta}_{k+1}
\end{equation}
Since this equation is actually a linear truncation of a Taylor series, we use iterative increments (indicated by $k$) to compensate \zcedit{for truncated nonlinearity}. In this equation, $\mathbf{r}_k \left( \in \mathbb{R} ^{N_{mod} \times \left( N_{sen} +1 \right)} \right)$ merges the normalized residue of modal eigenvalues, e.g., $\beta_\Lambda (\boldsymbol{\Lambda}_{Meas}-\boldsymbol{\Lambda}_{FEM})$, and the residue of the mass-normalized modal shapes, e.g., $\beta_\Phi (\boldsymbol{\Phi}_{Meas}-\boldsymbol{\Phi}_{FEM})$. $N_{mod}$ is the number of modes and $N_{sen}$ is the number of sensors. $\beta_\Lambda$ and $\beta_\Phi$ are normalization coefficients. The subscript ``Meas'' stands for \zcedit{measurements} and ``FEM'' denotes \zcedit{finite element model}. $\mathbf{S}_k \left( \in \mathbb{R} ^{N_{mod}\left( N_{sen} +1 \right) \times N_{ele}} \right)$ is the Jacobian matrix of the undamped FEM eigenvalues/shapes with respect to parameter $\boldsymbol{\theta}_k$ \cite{Fox}. \zcedit{Note that the sensitivity analysis of damped modal quantities can add more value to practical applications, however, here we aim to present the prototypical framework herein and leave that for the future work.} The last term $\Delta\boldsymbol{\theta}_k$ is the parameter increment. Hence, the identified stiffness parameters are the sum of the iterative increments, namely, $\boldsymbol{\theta} = \sum_{k} \Delta \boldsymbol{\theta}_k$.

\subsection{Model Updating: $\ell_2$ Bayesian Learning}
To update the parameters, we formulate the the ill-posed sensitivity equation (see Eq. (\ref{eq:SensitivityEquation})) in the context of hierarchical Bayesian inference. Essentially, solving the sensitivity equation is equivalent to finding the \zcedit{maximum \emph{a posteriori} (MAP)} estimate \cite{Sun2015b}, where the posterior \zcedit{probability density function} (PDF) of the unknown parameters, viz., $p \left( \Delta\boldsymbol{\theta}_{k+1}, \sigma_{k+1}^2, \alpha_{k+1} | \mathbf{r}_{k} \right)$, can be described as \cite{Chen}
\begin{equation} \label{eq:MAP}
\begin{aligned}
& p \left( \Delta\boldsymbol{\theta}_{k+1}, \sigma_{k+1}^2, \alpha_{k+1} | \mathbf{r}_{k} \right) \\
& \propto {} p \left( \mathbf{r}_{k} |  \Delta\boldsymbol{\theta}_{k+1}, \sigma_{k+1}^2 \right)
p \left( \Delta\boldsymbol{\theta}_{k+1} | \alpha_{k+1} \right) p \left( \sigma_{k+1}^2 \right) p \left( \alpha_{k+1} \right)
\end{aligned}
\end{equation}
where $p \left( \mathbf{r}_{k} |  \Delta\boldsymbol{\theta}_{k+1}, \sigma_{k+1}^2 \right)$ is the likelihood function which can be represented by the multivariate normal distribution $\mathcal{N}\left( \mathbf{S}_k \Delta\boldsymbol{\theta}_{k+1}-\mathbf{r}_{k}, \sigma_{k+1}^2 \mathbf{I} \right)$; $\sigma_{k+1}^2$ is the variance of the modeling error; $p \left( \Delta\boldsymbol{\theta}_{k+1} | \alpha_{k+1} \right)$ is the prior distribution of $\theta_{k+1}$ that follows the multivariate normal distribution $N \left( \mathbf{0}, \alpha_{k+1} \mathbf{I} \right)$ with the \zcedit{variance $\alpha_{k+1}$}; $p \left( \sigma_{k+1}^2 \right)$ and $p \left( \alpha_{k+1} \right)$ are the hyper-prior distributions of $\sigma_{k+1}^2$ and $\alpha_{k+1}$ following inverse Gamma distributions, namely, $IG \left( a_0, b_0 \right)$ and $IG \left( a_1, b_1 \right)$ respectively. Here, $\{a_0, a_1, b_0, b_1\}$ are user-defined hyperparameters \zcedit{that can be simply determined by magnitudes of other variables.}

To maximize the joint posterior PDF shown in Eq. (\ref{eq:MAP}), we firstly derive \zcedit{its closed form.} Then, we take \zcedit{partial derivatives of the joint posterior with respect to $\Delta\boldsymbol{\theta}$, $\sigma^2$ and $\alpha$, respectively} and set these derivatives to zeros, resulting in the following set of equations \cite{Sun2015b,Yan2017,Yan2019,Chen}
{\setlength{\mathindent}{0cm}
\begin{subequations}
\begin{align}
& \Delta\boldsymbol{\theta}_{k+1} = \left( \sum_{n=1}^{N_{ob}} (\mathbf{S}_k^n)^\top\mathbf{S}_k^n + \frac{\sigma_{k+1}^2}{\alpha_{k+1}} \mathbf{I} \right)^{-1} \sum_{n=1}^{N_{ob}} (\mathbf{S}_k^n)^\top \mathbf{r}_k^n \\
& \sigma_{k+1}^2 = m^{-1} \left( \sum_{n=1}^{N_{ob}} \| \mathbf{S}_k^n \Delta\boldsymbol{\theta}_{k+1} - \mathbf{r}_k^n \|^2 + 2b_0 \right) \\
& \alpha_{k+1} = \left[ N_{ele}/2 + \left( a_1 +1 \right) \right]^{-1} \left( \| \Delta\boldsymbol{\theta}_{k+1} \|^2 + 2b_1 \right) 
\end{align}
\end{subequations}}%
\noindent where $m = N_{ob}N_{mod}(N_{sen} +1)+ 2(a_0 + 1)$ and $N_{ob}$ is the number of independent data sets (observations). By sequentially updating $\Delta\boldsymbol{\theta}_{k+1}, \sigma_{k+1}^2$ and $\alpha_{k+1}$, we can get their optimum after \zcedit{a few} iterations. This essentially forms an automatic Bayesian learning process \cite{Sun2015b, Sun2015}.

\zcedit{Note that} the MAP only provides a deterministic prediction of $\Delta\boldsymbol{\theta}_{k+1}, \sigma_{k+1}^2$ and $\alpha_{k+1}$. \zcedit{A probabilistic estimation for quantifying how reliable the MAP estimate is can be achieved by approximating} the posterior $p\left( \Delta\boldsymbol{\theta}_{k+1} | \mathbf{r}_{k} \right)$ with a multivariate Gaussian distribution. \zcedit{We compute the Hessian matrix of the negative logarithmic form of the posterior PDF and derive} the covariance matrix \cite{Yuen,Sun2015c}: 
\begin{equation} \label{eq:HessianCovariance}
\begin{aligned}
\mathcal{H}\left(\Delta\boldsymbol{\theta}_{k+1}\right) & = \frac{\partial^2\mathcal{J}\left(\Delta\boldsymbol{\theta}_{k+1}, \sigma_{k+1}^2, \alpha_{k+1}\right)}{\partial\left(\Delta\boldsymbol{\theta}_{k+1}\right)^2} \\
& = N_{ob}\boldsymbol{\Sigma}^{-1}_{\Delta\boldsymbol{\theta}_{k+1}}
\end{aligned}
\end{equation}
where $\mathcal{H}(\Delta\boldsymbol{\theta}_{k+1})$ is the Hessian matrix of $\Delta\boldsymbol{\theta}_{k+1}$; $\mathcal{J} = -\ln{p \left( \Delta\boldsymbol{\theta}_{k+1}, \sigma_{k+1}^2, \alpha_{k+1} | \mathbf{r}_{k} \right)}$ is the \zcedit{loss} function; and $\boldsymbol{\Sigma}^{-1}_{\Delta\boldsymbol{\theta}_{k+1}}$ denotes the covariance matrix of the posterior $p\left( \Delta\boldsymbol{\theta}_{k+1} | \mathbf{r}_{k} \right)$. Strictly speaking, if we use conjugate prior, the joint posterior $p \left( \Delta\boldsymbol{\theta}_{k+1}, \sigma_{k+1}^2, \alpha_{k+1} | \mathbf{r}_{k} \right)$ will be normal-inverse-Wishart distribution and its marginal posterior $p\left( \mathbf{S}_k \Delta\boldsymbol{\theta}_{k+1} | \mathbf{r}_{k} \right)$ follows a student's t distribution\cite{Bishop}. Besides, as the Central Limit Theorem \cite{Rosenblatt} states in most cases, as the number of independent observations goes up the aggregated distribution will approach a Gaussian distribution. Therefore, modeling the posterior distribution $p\left( \Delta\boldsymbol{\theta}_{k+1} | \mathbf{r}_{k} \right)$ by a Gaussian distribution \zcedit{can be justified}.

Now that we have a joint posterior distribution of the stiffness parameters, we aim to further obtain the marginal posterior distribution of each parameter, e.g., $p(\theta^{i} | \mathbf{r})$. We design a Monte Carlo sampling strategy to obtain $p(\theta^{i} | \mathbf{r})$ based on the procedure as follows: firstly, we accumulate all increments to obtain the MAP estimate $\hat{\boldsymbol{\theta}} = \sum_{k} \Delta \boldsymbol{\theta}_k$; secondly, we compute the aggregated variance $\hat{\boldsymbol{\Sigma}}_\theta = \sum_{k} \boldsymbol{\Sigma}_{\Delta\boldsymbol{\theta}_{k+1}}$; \zcedit{then}, we draw a vast number of samples for each parameter $\theta^{i}$ using the joint normal distribution $\mathcal{N}(\hat{\boldsymbol{\theta}},  \hat{\boldsymbol{\Sigma}}_\theta )$; \zcedit{lastly}, we fit a Gaussian distribution to the samples and obtain the posterior distribution of each parameter, e.g., $p(\hat{\theta}^{i} | \mathbf{r})$. Given the quantified mean and variance of the stiffness parameters, \zcedit{we can evaluate the uncertainty}.

\subsection{Sparse Damage Identification: Sequential Threshold Least Squares Regression}
\subsubsection{\zcedit{Sequential Threshold Least Squares Regression.}}
\zcedit{Structural damage often occurs locally at a few locations posing a sparse distribution nature.} In the context of stiffness parameter variation identification, \zcedit{sparsity is present in such cases (i.e.}, $\Delta\boldsymbol{\theta}_{k+1}$ is sparse) where sparse representation should be applied. Therefore, the sensitivity analysis as shown in Eq. (\ref{eq:SensitivityEquation}) can be cast as an ideal $\ell_0$ optimization problem, expressed as
\begin{equation} \label{eq:L0}
\Delta\hat{\boldsymbol{\theta}}_{k+1} = \underset{\Delta\boldsymbol{\theta}_{k+1}}{\arg \min} \|\Delta\boldsymbol{\theta}_{k+1}\|_0 \quad \text{s.t.} \quad \mathbf{r}_k = \mathbf{S}_k \Delta \boldsymbol{\theta}_{k+1}
\end{equation}
where the $\ell_0$ norm $\|\Delta\boldsymbol{\theta}_{k+1}\|_0$ counts the non-zero values in $\Delta\boldsymbol{\theta}_{k+1}$. Despite its ideal characteristics for sparse regression, its non-differentiablity makes its optimization an NP-hard problem, possessing extremely computational complexity. The Sequential Threshold Least Squares (STLS) regression \cite{Zhang2019,Rudy} provides an elegant alternative to sparsely solve the $\ell_0$ sensitivity equation in Eq. (\ref{eq:L0}) in a quasi-$\ell_0$ manner. The core concept of STLS is to turn the $\ell_0$ sparse representation into a series of \zcedit{least-squares} regression processes with hard thresholds. The STLS regression approach has been proven more effective and more accurate than classical $\ell_1$ method in regard to promoting sparsity \cite{Zhang2019}. Herein, we turn Eq. (\ref{eq:L0}) to the STLS regression, written as:
\begin{equation} \label{eq:STLS}
\Delta \hat{\boldsymbol{\theta}}^{j+1}_{k+1} = \underset{\text{supp} (\Delta \boldsymbol{\theta}_{k+1}) \subseteq \mathcal{B}^j_{k+1}}{\arg\min} \|\mathbf{r}_{k} - \mathbf{S}_{k} \Delta \boldsymbol{\theta}_{k+1}\|_2
\end{equation}
where
\begin{equation}
\mathcal{B}^j_{k+1} = \left \{ 1 \leq i \leq N_{ele}: | \Delta \hat{\theta}^j_{i(k+1)}|\geq \lambda \right \}
\end{equation}
Note that the STLS algorithm should be run in each sensitivity iteration $k$. The general workflow of Eq. (\ref{eq:STLS}) is that, in the $(j+1)$th STLS iteration, $\Delta \hat{\boldsymbol{\theta}}^{j+1}_{k+1}$ is obtained by the least squares solution of $\mathbf{r}_k = \mathbf{S}_k \Delta \boldsymbol{\theta}_{k+1}$, meanwhile $\Delta \boldsymbol{\theta}_{k+1}$ also needs to be a support set of $\mathcal{B}^j_{k+1}$ from the last iteration. Note that $\mathcal{B}^j_{k+1}$ only includes those entries $\Delta \hat{\theta}^j_{i(k+1)}$ either greater than or equal to an adaptive threshold $\lambda$ and the rest entries are excluded and set to zeros.

\begin{algorithm}[t!]
\setstretch{1.1}
\caption{\small{Sequential Threshold Least Squares (STLS) regression: $[\Delta \hat{\boldsymbol{\theta}}, L] = \texttt{STLS}\left(\mathbf{r}, \mathbf{S}, \lambda\right)$}}
\label{alg:STLS}
\small
\begin{algorithmic}[1]

\STATE {\bfseries Input:} Sensitivity residue $\mathbf{r}$, Jacobian matrix $\mathbf{S}$ and sparsity threshold $\lambda$.

\STATE {\bfseries Initialize $\Delta \hat{\boldsymbol{\theta}}^j$:} Estimate $\Delta \hat{\boldsymbol{\theta}}^0$ from LASSO using cross validation $\Delta \hat{\boldsymbol{\theta}}^{0} = {\arg\min} \| \mathbf{r} - \mathbf{S} \Delta \boldsymbol{\theta}\|_2^2 + \eta \| \Delta \boldsymbol{\theta} \|_1$, where $\eta$ is optimally selected from a geometric series.

\STATE {\bfseries Initialize loss $L^{best}$:} \zcedit{Compute the initial best loss $L^{best} = \big\| \mathbf{r} - \mathbf{S} \Delta \hat{\boldsymbol{\theta}}^{0} \big\|_2 + 0.001 \text{cond} (\mathbf{S}) \big\| \Delta \hat{\boldsymbol{\theta}}^{0}\big\|_0 $.} 

\WHILE {$j \leq 10$}
	
    \STATE Threshold entries of $\Delta \hat{\boldsymbol{\theta}}^{j-1}$ by $\lambda$: $B^{j-1} = \{ 1 \leq i \leq N_{ele}: | \Delta \hat{\theta}^{j-1}_i|\geq \lambda \}$ and $A^{j-1} = \{ 1 \leq i \leq N_{ele}: | \Delta \hat{\theta}^{j-1}_i| < \lambda \}$. 
    
    \STATE Enforce zeros to smaller entries $\Delta \hat{\boldsymbol{\theta}}^{j-1} \left(A^{j-1} \right) = 0$. Update remaining non-zero entries (a.k.a $B^{j-1}$) by least squares regression.
    
    \STATE Combine $B^{j-1}$ and $A^{j-1}$ to form $\Delta \hat{\boldsymbol{\theta}}^{j}$. 
    
    \STATE Compute the new loss $L^{j}$.
    
    \IF {$L^j < L^{best}$ and $\| \Delta \hat{\boldsymbol{\theta}}^{j} \|_0 \neq 0$}
        \STATE $L^{best} = L^j$
        \STATE $\Delta \hat{\boldsymbol{\theta}}^* = \Delta \hat{\boldsymbol{\theta}}^{j}$
    \ELSE
        \STATE Break the \textbf{while} loop.
    \ENDIF
    \STATE $j = j+1$.
\ENDWHILE

\STATE {\bfseries Output:} $\Delta \hat{\boldsymbol{\theta}} = \Delta \hat{\boldsymbol{\theta}}^*, L = L^{best}$.

\end{algorithmic} 
\end{algorithm}

\begin{algorithm}[t!]
\setstretch{1.1}
\caption{\small{Bayesian hyperparameter Optimization: $\big[\lambda_{best}, \Delta \hat{\boldsymbol{\theta}}^{best}\big] = \texttt{BayesOpt}\left( \mathbf{r}, \mathbf{S}, \lambda_{min}, \lambda_{max} \right)$}}
\label{alg:BayesOpt}
\small
\begin{algorithmic}[1]
\STATE {\bfseries Input:} Sensitivity residue $\mathbf{r}$, Jacobian matrix $\mathbf{S}$ and the low bound and the high bound for sparsity threshold $\lambda_{min}$ and $\lambda_{max}$. In this case, $\lambda_{min}=0.01$ and $\lambda_{max}=1$.

\STATE {\bfseries \zcedit{Initialize $\lambda_{best}$}:} \zcedit{Evaluate $[\sim, L_i] = \texttt{STLS}\left(\mathbf{r}, \mathbf{S}, \lambda_i\right)$ for four randomly sampled points ($i = 1,2,3,4$) from $[\lambda_{min}, \lambda_{max}]$. Determine $\lambda_{best}$ that has the smallest $L_i$.}

\STATE {\bfseries \zcedit{Initialize the Gaussian Process (GP) model:}} \zcedit{Model the \zcedit{loss} function $L(\lambda)$ as a GP model $f(\lambda, \kappa, \sigma^2_{noise})$ with mean $\mu(\lambda; \kappa)$, the ARD Matérn 5/2 covariance kernel \cite{Rasmussen} $k(\lambda, \lambda'; \kappa)$ and Gaussian noise with variance $\sigma^2_{noise}$}. 

\WHILE {$j \leq 30$}
    \STATE \zcedit{Update the GP model by computing the posterior distribution $Q(\kappa, \sigma^2_{noise}| \lambda_i, L_i \textrm{ for } i = 1,2,3,...)$.}
    \STATE \zcedit{Find the next $\lambda_{new}$ by maximizing the acquisition function $a \left( \lambda \right) = E_Q \left[ \max \left( 0, L (\lambda_{best}) - f(\lambda, \kappa, \sigma^2_{noise}) \right) \right]$.}
    \STATE \zcedit{Evaluate $L_{new}$ for the new sampling point $\lambda_{new}$}
    \IF {\zcedit{$L_{new} < L_{best}$}}
        \STATE \zcedit{$\lambda_{best} = \lambda_{new}$}
    \ELSE
        \STATE \zcedit{Add $(\lambda_{new},L_{new})$ into $(\lambda_i,L_i).$}
    \ENDIF
    \STATE $j = j + 1$.
\ENDWHILE

\STATE {\bfseries Output:} \zcedit{$\lambda_{best}$ and $\big[\Delta \hat{\boldsymbol{\theta}}^{best}, \sim\big] = \texttt{STLS}\left(\mathbf{r}, \mathbf{S}, \lambda_{best}\right)$}.

\end{algorithmic} 
\end{algorithm}

\subsubsection{\zcedit{Bayesian Hyperparameter Optimization.}}
Obviously, the threshold $\lambda$ is critical for controlling the sparsity \zcedit{and attaining regression accuracy. The selection of $\lambda$ is dependent on the linear system in Eq. (\ref{eq:L0}). However, $\mathbf{r}_k, \mathbf{S}_k$ and $\Delta \boldsymbol{\theta}_{k+1}$ change in every sensitivity iteration $k$. Thus, manually setting a constant $\lambda$ may not be a wise choice. Therefore, we propose to leverage Bayesian hyperparameter optimization \cite{Snoek} to heuristically determine $\lambda$ in each sensitivity iteration. First of all, we define a loss function $L$ for depicting the balance role of $\lambda$:
\begin{equation} \label{eq:LossforSTLS}
L(\lambda) = \| \mathbf{r}_k - \mathbf{S}_k \Delta \boldsymbol{\theta}_{k+1} (\lambda) \|_2 + \delta \gamma_k \| \Delta \boldsymbol{\theta}_{k+1} (\lambda) \|_0 
\end{equation}
\noindent where $\gamma_k = \text{cond} (\mathbf{S}_k)$ and $\text{cond} (\cdot)$ computes the condition number, while the coefficient $\delta$ is to make the two summation terms at about the same magnitude (e.g., $\delta=0.001$). We can see that $L$ and $\lambda$ have an implicit relationship. To proxy this relationship, we build a Gaussian Process (GP) model that is updated by past evaluations of the loss function in Eq. (\ref{eq:LossforSTLS}). 

We denote the initial GP model by $f(\lambda, \kappa, \sigma^2_{noise})$ with mean $\mu(\lambda; \kappa)$, the ARD Matérn 5/2 covariance kernel \cite{Rasmussen} $k(\lambda, \lambda'; \kappa)$ and Gaussian noise with variance $\sigma^2_{noise}$ ($\kappa$ is a hyperparameter of the kernel). Then, we randomly choose a set of initial values $\lambda_i~(i = 1,2,3,...)$ from the hyperparameter's bounds and update the GP model by calculating the posterior distribution $Q(\kappa, \sigma^2_{noise}|\lambda_i, L_i \textrm{ for } i = 1,2,3,...)$. By applying an acquisition function $a(\lambda)$ to the GP model, which in our case is the expected improvement (see Eq. (\ref{eq:ExpectedImprovement})), we can locate a new value for $\lambda$. After sequentially updating the GP model, maximizing the acquisition function and evaluating the loss function $L(\lambda)$ with a new $\lambda$ for many iterations, we can approach the optimal value of $\lambda$. Note that the expected improvement is defined as
\begin{equation} \label{eq:ExpectedImprovement}
a \left( \lambda \right) = E_Q \left[ \max \left( 0, L (\lambda_{best}) - f(\lambda, \kappa, \sigma^2_{noise}) \right) \right]
\end{equation}
where $\lambda_{best}$ is the current best option.} It has been shown that Bayesian optimization can lead to faster convergence than grid or random search \cite{Snoek}, due to the ability to prioritize the search area. 

\subsubsection{\zcedit{Least Absolute Shrinkage and Selection Operator.}}
\zcedit{It} is worth mentioning that a desirable performance of STLS requires a good initialization of $\Delta \hat{\boldsymbol{\theta}}^{0}_{k+1}$, which can be estimated by Least Absolute Shrinkage and Selection Operator (LASSO) \cite{Tibshirani} as 
\begin{equation} \label{eq:Lasso}
\Delta \hat{\boldsymbol{\theta}}^{0}_{k+1} = \underset{\Delta \boldsymbol{\theta}_{k+1}}{\arg\min} \| \mathbf{r}_k - \mathbf{S}_k \Delta \boldsymbol{\theta}_{k+1}\|_2^2 + \eta \| \Delta \boldsymbol{\theta}_{k+1} \|_1
\end{equation}
where $\eta$ is a pivotal regularization coefficient controlling how many entries are close to to zeros. To choose a suitable $\eta$, we guess a geometric series of 100 $\eta$'s, with the last one (the largest one) as large as possible \zcedit{such that it refrains $\boldsymbol{\theta}_{k+1}$ from becoming all zeros} and the first one (the smallest one) having a $1\times10^{-4}$ ratio with respect to the last one. Then, we select the $\eta$ \zcedit{whose corresponding loss in Eq. (\ref{eq:Lasso})} ranked the second smallest among all the 100 losses. The reason for not choosing $\eta$ with the smallest loss is \zcedit{to alleviate overfitting to noise. Furthermore,} we cross-validate the LASSO model by randomly splitting $\mathbf{r}$ and $\mathbf{S}$ into $n$ portions, using $n-1$ portions as training data when the regression loss is computed on the last portion, repeating this process $n$ times, and averaging all the $n$ models \zcedit{to get an average estimation of $\eta$}. 

More details of how to implement the proposed STLS algorithm are presented in the pseudo-code (Algorithm \ref{alg:STLS} and Algorithm \ref{alg:BayesOpt}). An important note is that the entire algorithm is designed to run in each sensitivity iteration. Even though STLS is a point-estimate method by its definition, we can still build confidence intervals by doing more tests and computing the statistical variance.

\section{Numerical and Experimental Examples} \label{Example}
In this section, we presents three case studies to validate our proposed framework. The first two are numerical simulations of a 10-story shear-type model and a 31-member truss structure. The third one is a shake-table test of an 8-DOF steel frame. The proposed computational framework was coded in MATLAB \cite{Moore} on a standard workstation with 10 Intel i9 CPU cores and 64GB memory.

\subsection{Numerical Example: A 10-Story Shear-Type Model}
\begin{figure}[t]
\centering
\includegraphics[width=0.3\textwidth]{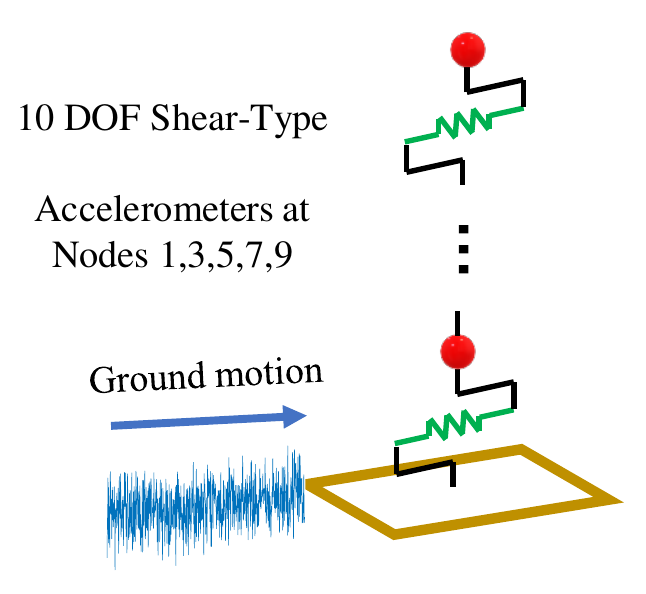}
\caption{A 10-story shear type model. Element stiffness is 176.729 MN/m and node mass is 100 ton. Accelerations from the odd nodes are known. 5 different white noise excites structural vibration.}
\label{Fig:10DOF}
\end{figure}

This proof-of-concept example is a 10-story shear-type model (see Figure \ref{Fig:10DOF}), whose nominal inter-layer stiffness and mass are 176.729 MN/m and 100 ton \cite{Yuen2006, Chen}. The first two damping ratios are 2\%. The element stiffness parameters \zcedit{in the actual system} are assumed to fluctuate between --20\% and 20\%. For the purpose of damage identification, we assume that \zcedit{28\%} and 33\% stiffness reduction are present in the first and the third stories respectively (count from bottom to top).

Five accelerometers are installed on the \zcedit{odd floors} and a duration of \zcedit{10-min response} under white-noise ground motion excitation is recorded for both undamaged and damaged cases. Five monitoring tests are conducted and all measurements are polluted by noise, whose Root Mean Square (RMS) is 10\% of \zcedit{that of} the clean signal. \zcedit{The first three modal frequencies and shapes are extracted by the Observer/Kalman filter IDentidication (OKID) \cite{Juang} followed by the Eigensystem Realization Algorithm (ERA) \cite{Juang1985}, namely, OKID/ERA, which computes the Markov parameters of an observer (e.g. the Kalman filter) and further identifies the state-space model and the modal parameters used for model updating and damage identification \cite{Lus}.}

\begin{figure}[h]
\centering
\includegraphics[width=0.35\textwidth]{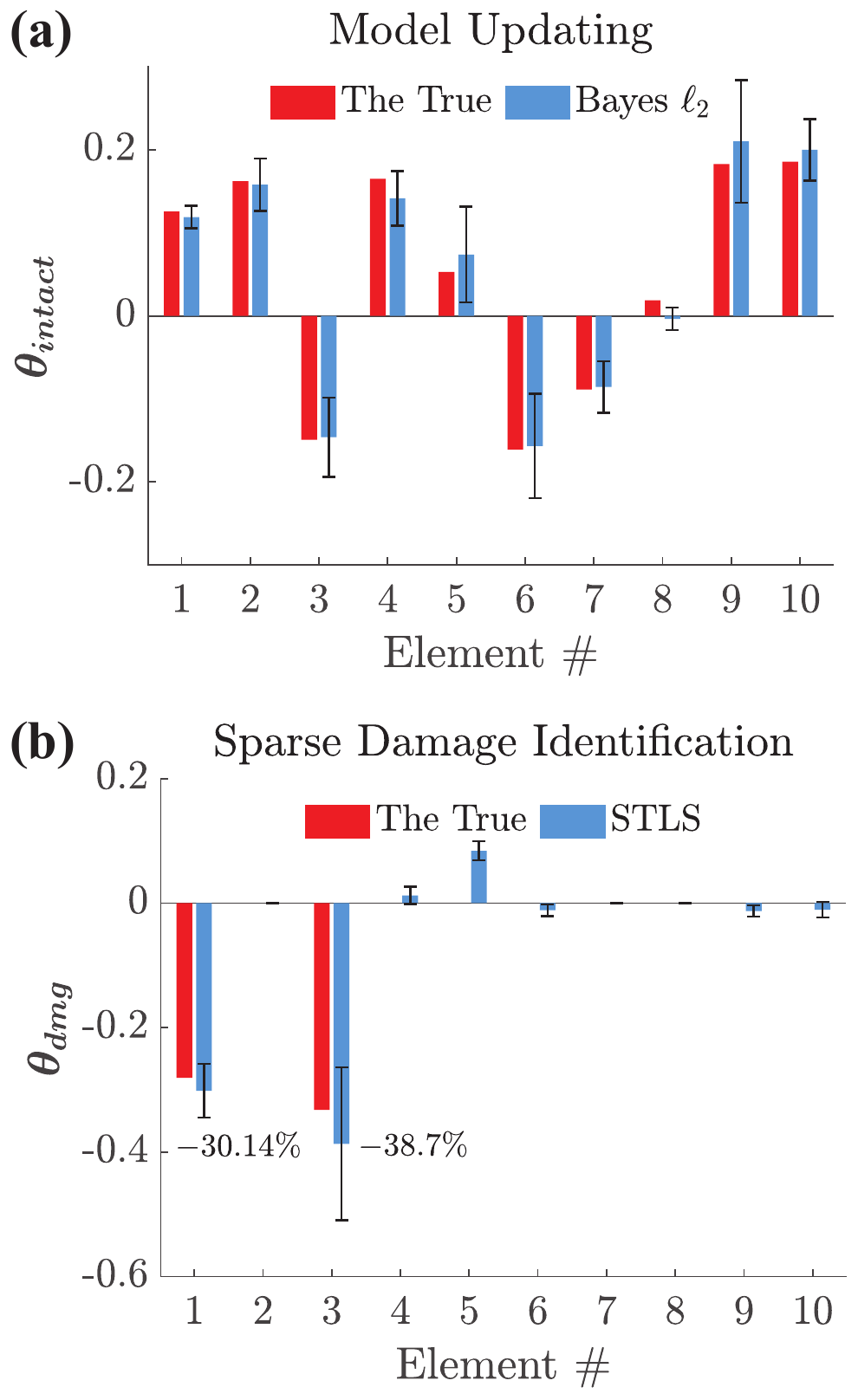}
\caption{Stiffness variation coefficients in model updating and sparse damage identification under 10\% noise for the 10-DOF shear-type structure: (a) the prediction is almost in accordance with the ground truth with error bars showing 95\% confidence interval; (b) even though there are very minor false positives(the highest is less than 10\%), \zcedit{the two most important components are very clear}.}
\label{Fig:10DOF_MUDI}
\end{figure}

\begin{figure*}[h]
\centering
\includegraphics[width=0.88\textwidth]{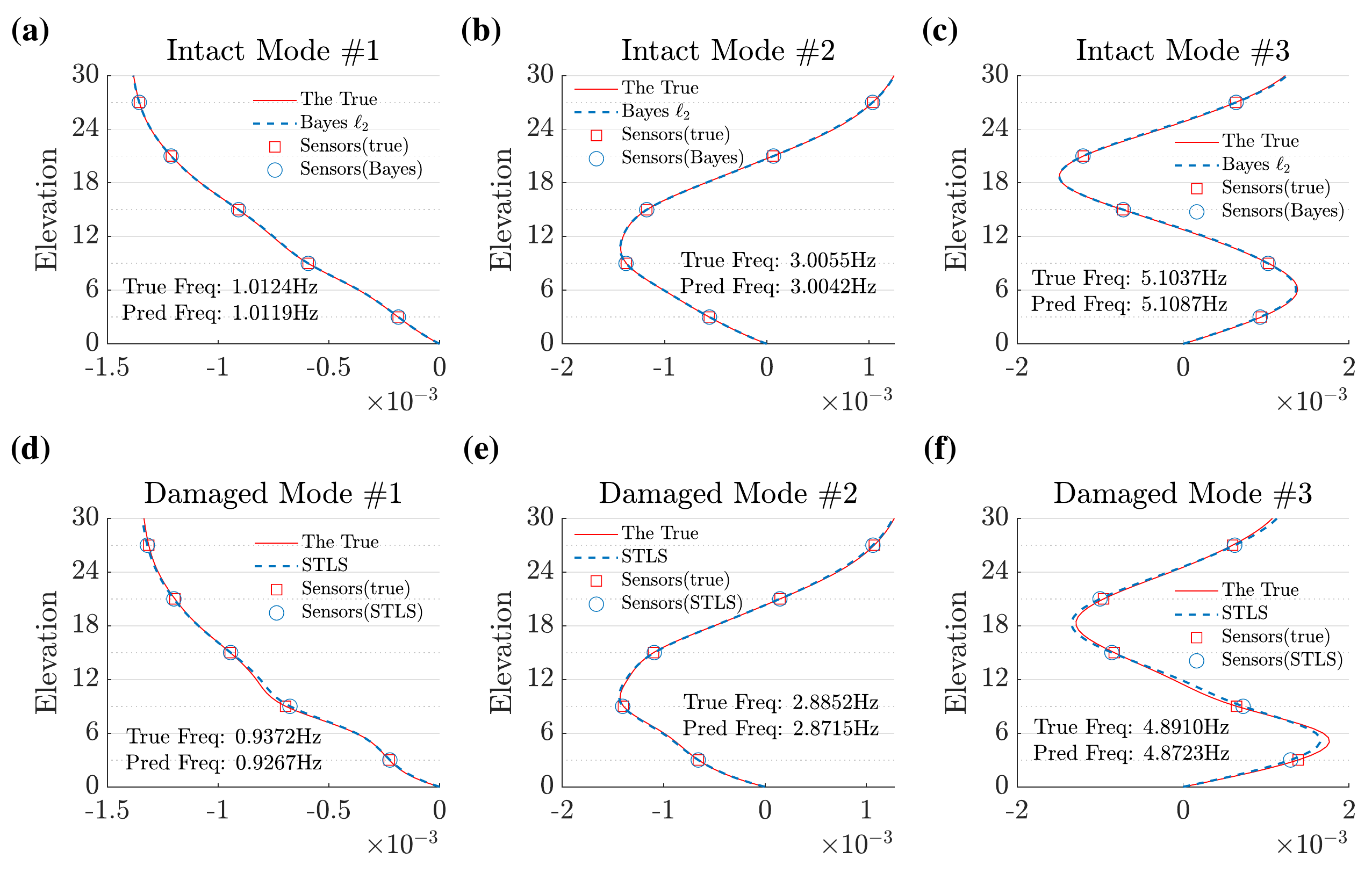}
\caption{Comparison between the first three modes after model updating/damage identification for the 10-DOF shear-type structure: (a$\sim$c) the upper panel displays the predicted and the true mode frequencies and shapes in the intact model, where the frequency has an impressive average error of 0.08\%; (d$\sim$e) the lower panel illustrates the mode frequencies and shapes in the damaged model, \zcedit{where the frequency's average divergence is still satisfyingly small at 0.43\%.}}
\label{Fig:10DOF_Shapes}
\end{figure*}

\begin{figure*}[h]
\centering
\includegraphics[width=0.9\textwidth]{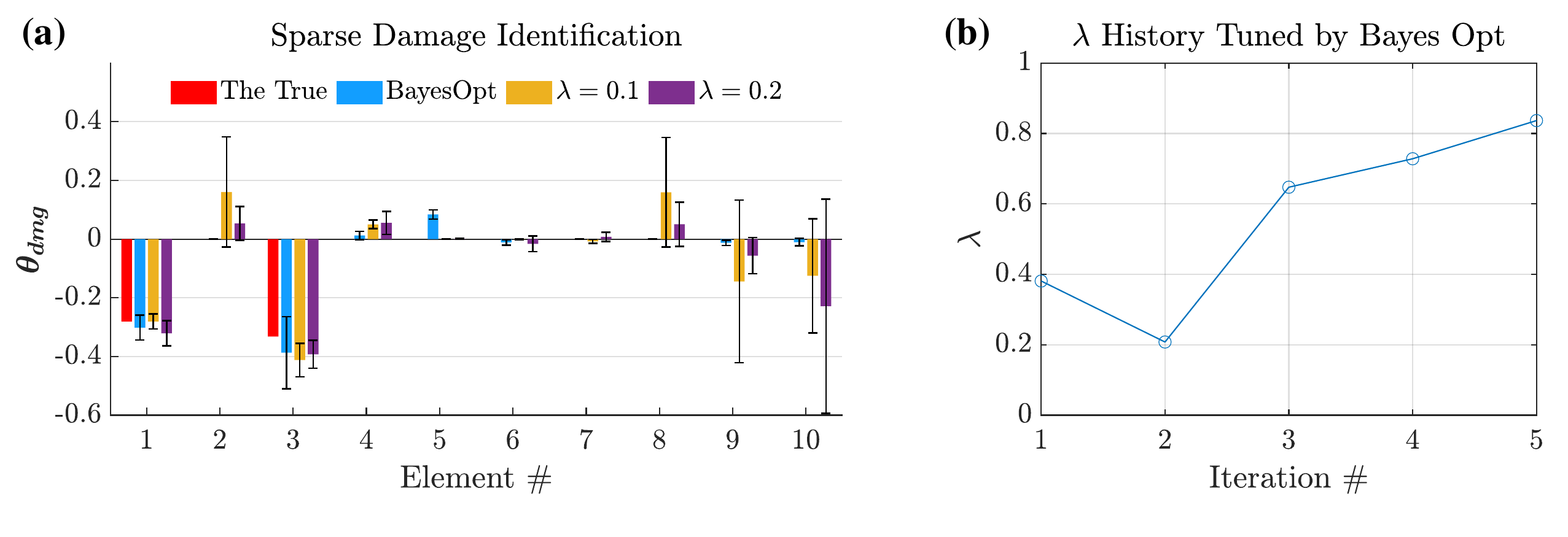}
\caption{\zcedit{Effects of Bayesian optimization in the 10DOF shear-type example: (a) Damage identification results when using Bayesian optimization to determine $\lambda$,or manually setting $\lambda=$0.1 or 0.2 throughout the sensitivity analysis; (b) $\lambda$ values determined by Bayesian optimization in each sensitivity iteration.}}
\label{Fig:BayesOptCompare}
\end{figure*}

The updated stiffness variation parameters of the intact model, e.g., $\boldsymbol{\theta}_{intact}$, are shown in Figure \ref{Fig:10DOF_MUDI}(a). It can be seen that the $\ell_2$ Bayesian learning approach can produce accurate identification of the stiffness variation parameters \zcedit{that has} a high correlation with the ground truth. The 95\% confidence intervals (computed from variance of the marginal posterior distributions) can well cover the ground truth. While for damage identification (see Figure \ref{Fig:10DOF_MUDI}(b)), the STLS method successfully identifies the stiffness reduction occurred in the first and the third stories. Minor false positives are also observed which might be due to inaccuracy of the updated intact model and/or noise pollution. Nevertheless, the overall performance of the proposed two-stage model updating and damage identification framework is \zcedit{satisfactory}. In addition, Figure \ref{Fig:10DOF_Shapes} depicts the updated modal quantities in comparison with the ground truth, which shows accurate prediction. \zcedit{We further analyze the effectiveness of Bayesian hyperparameter optimization. A comparison between Bayesian optimization and the case of manually selected constant $\lambda$ is illustrated in Figure \ref{Fig:BayesOptCompare}. It can be seen that two tentative trials of $\lambda$ with 0.1 and 0.2 provide less satisfactory identification with multiple false positives and large uncertainties, especially for elements 8, 9 and 10. In contrast, Bayesian hyperparameter optimization heuristically determines $\lambda$ in each iteration of the sensitivity analysis (see Figure \ref{Fig:BayesOptCompare}(b)) leading to more accurate damage identification.}


\begin{figure*}[h]
\centering
\includegraphics[width=0.65\textwidth]{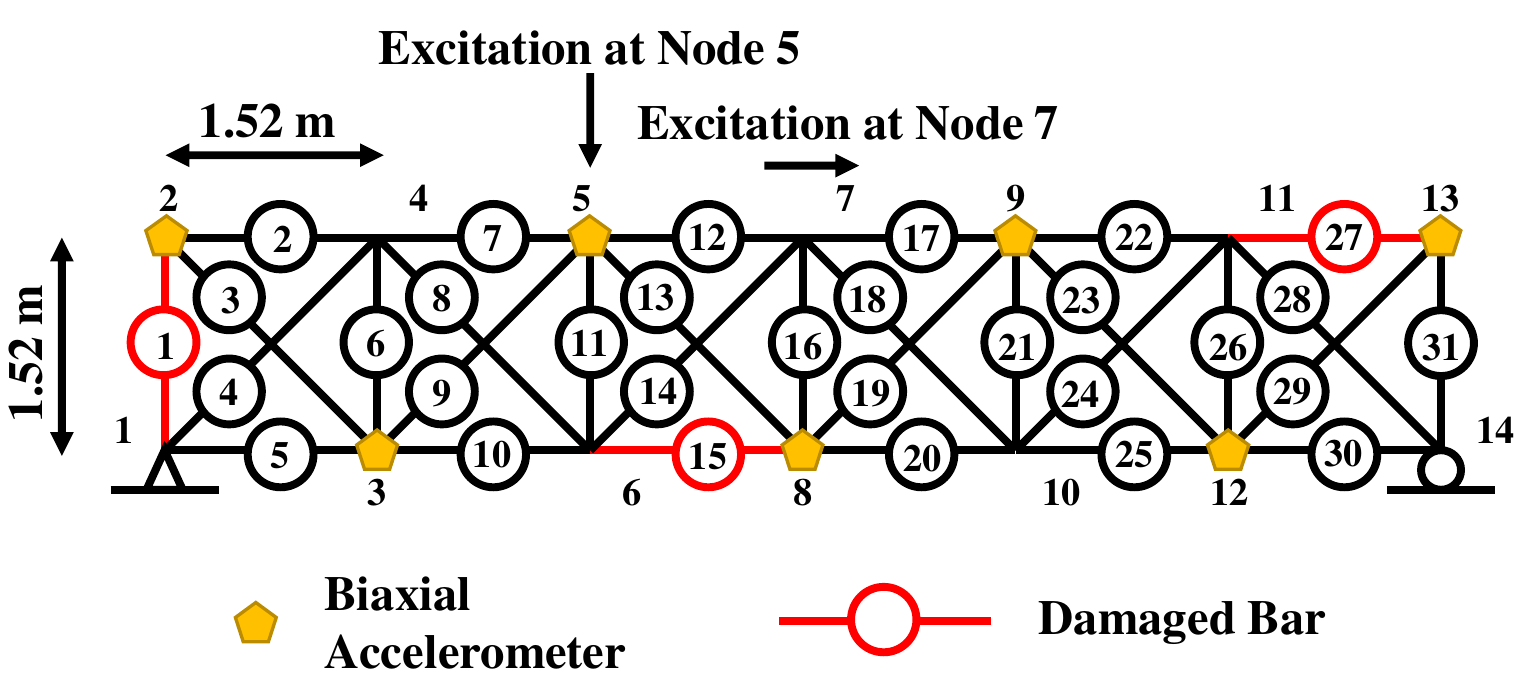}
\caption{Illustration of the 31-bar plane truss with sensor locations. White noise are imposed on the vertical direction and the horizontal direction of Node 5 and Node 7, respectively.}
\label{Fig:Truss}
\end{figure*}

\begin{figure*}[h]
\centering
\includegraphics[width=0.72\textwidth]{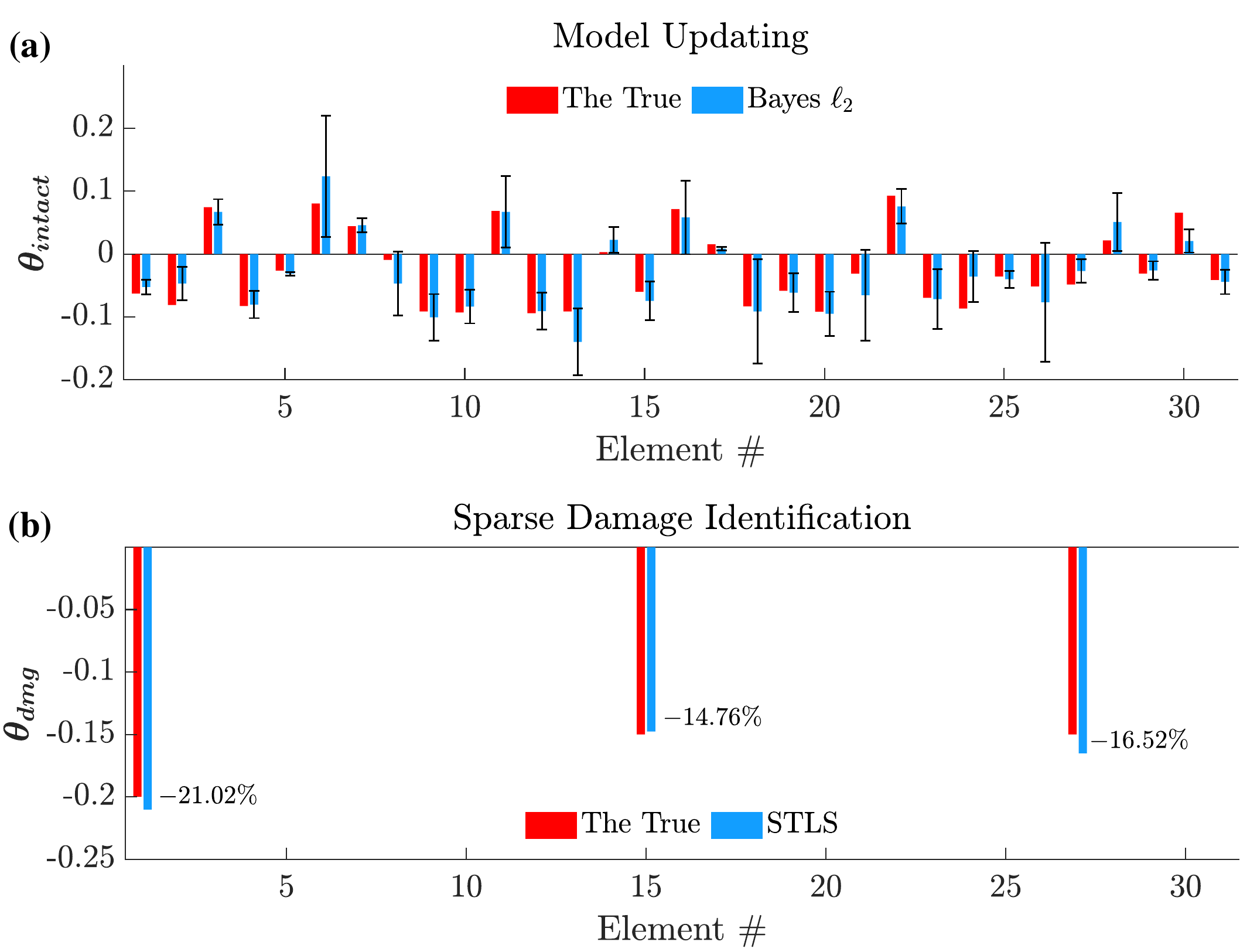}
\caption{Stiffness variation coefficient after model updating/sparse damage identification using one measurement with 10\% RMS noise for the truss structure: (a) the Bayesian posterior mean reliably estimates the majority of stiffness variation in the intact model, with a correlation coefficient of 0.93; (b) the STLS very accurately capture the sparsity pattern in the damaged stiffness, with a correlation coefficient close to 1.}
\label{Fig:Truss_MUDI}
\end{figure*}

\begin{figure*}[h]
\centering
\includegraphics[width=0.75\textwidth]{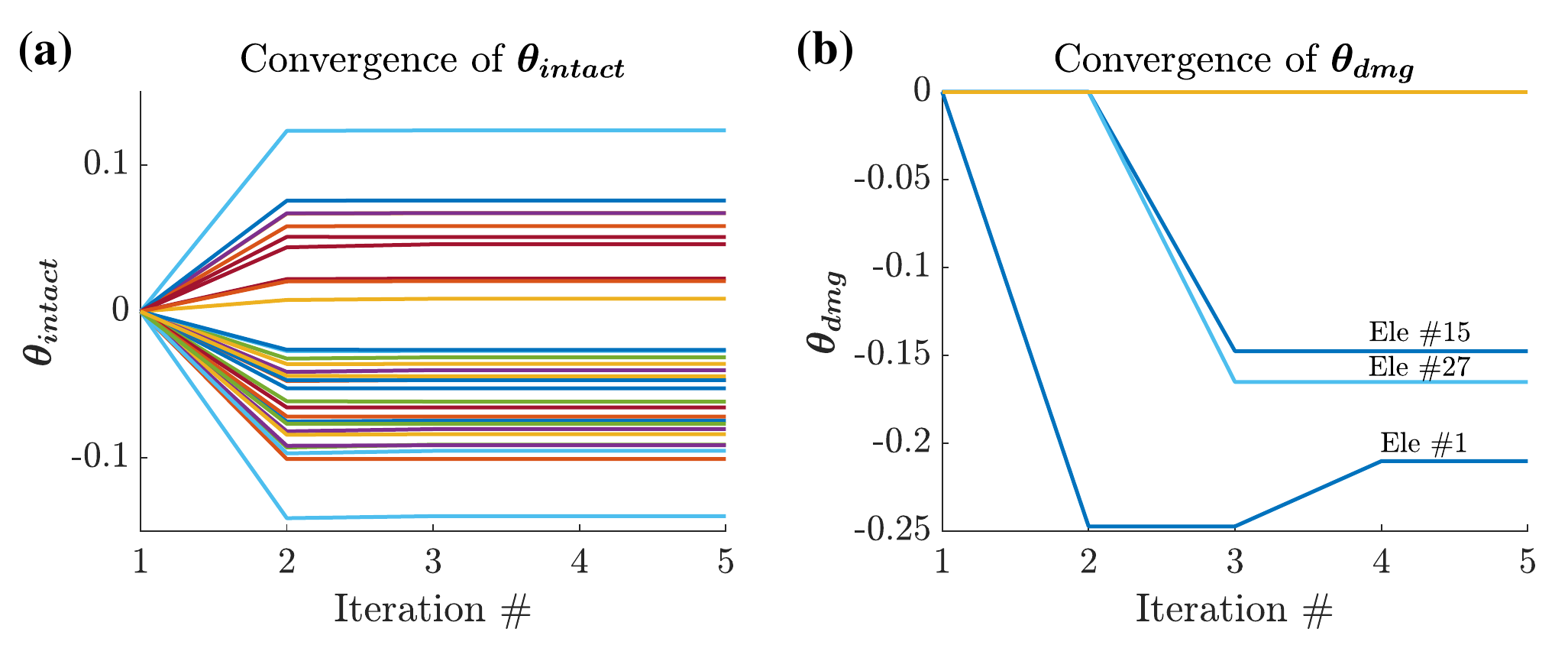}
\caption{Convergence of $\boldsymbol{\theta}$ in model updating and sparse damage identification for the truss structure. In this case, both $\boldsymbol{\theta}_{intact}$ and $\boldsymbol{\theta}_{dmg}$ stabilize in very few iterations.}
\label{Fig:Truss_CNVG}
\end{figure*}

\subsection{Numerical Example: A 31-Member Truss Structure}
Here we consider a large advertising stand, modeled as a 31-bar simply-supported truss structure \cite{Li, Chen} as shown in Figure \ref{Fig:Truss}, with its material/geometry properties given as follows: elastic modulus 70 GPa, cross-section 25 cm$^2$ and material density 2770 kg/m$^3$. The damping ratios for the first two dominant modes are 1\% and 2\%. To implement model updating, we assume there's a random variation in $[-10\%,10\%]$ for each element's stiffness. To showcase damage detection, the stiffness of Bar 1 reduces by 20\% while the stiffness reductions of Bars 15 and 27 \zcedit{are} 15\%. The structure is excited by a white noise force in the vertical direction of Node 5 and the horizontal direction of Node 7 at the same time as shown in Figure \ref{Fig:Truss}. In this example, we tend to explore our framework's performance using scarce and noisy data. In particular, biaxial acccelerometers are deploy at Nodes 2, 3, 5, 8, 9, 12 and 13, which record the structural response for 60 s at 1400 Hz. Only one set of measurement under 10\% RMS noise is recorded (for intact model updating and damage identification, respectively) and processed by OKID/ERA \cite{Juang,Juang1985} for modal extraction.

Figure \ref{Fig:Truss_MUDI} shows the result for both intact model updating and damage identification. It can be seen from Figure \ref{Fig:Truss_MUDI}(a) that the updated stiffness variation parameters in general matches well the ground truth with acceptable discrepancies (e.g., the correlation coefficient is 93\%). The large-level of noise and the limited number of sensors cause less satisfactory identification of parameters with small values. Nevertheless, most of the parameters have small variance showing reliable robustness. For those with large uncertainties, more tests and sensors may be helpful to reduce the bias. In the stage of damage identification, it is encouraging that the STLS successfully localizes the three damaged elements while accurately quantifying the damage extents. Since we only utilize one set of measurement, we cannot provide statistical mean and variance for the STLS result. Figure \ref{Fig:Truss_CNVG} shows the \zcedit{convergence} histories of (a) the proposed $\ell_2$ Bayesian learning for intact model updating and (b) the STLS regression for damage identification. It appears that both methods converge very quickly with only a few number of iterations, in which the convergence tolerance is set to be a relative error of $1\times10^{-6}$. 

\begin{figure*}[h]
\centering
\includegraphics[width=0.85\textwidth]{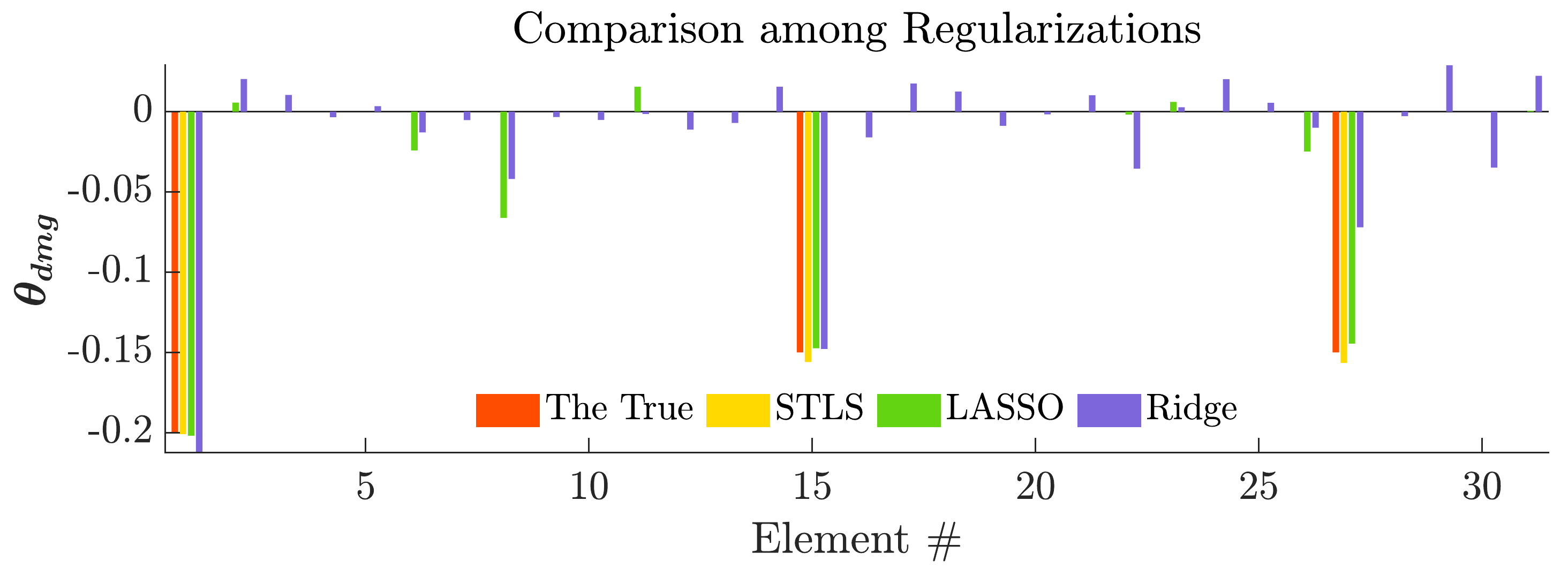}
\caption{Comparison among STLS, LASSO and ridge regularizations for sparse damage identification for the truss structure. The STLS is in close proximity to the true sparse pattern. LASSO manifests similar competence in spite of a handful of false positives. Ridge regression appears to have the smoothest result.}
\label{Fig:Truss_ReguComparison}
\end{figure*}

\begin{figure}[h]
\centering
\includegraphics[width=0.475\textwidth]{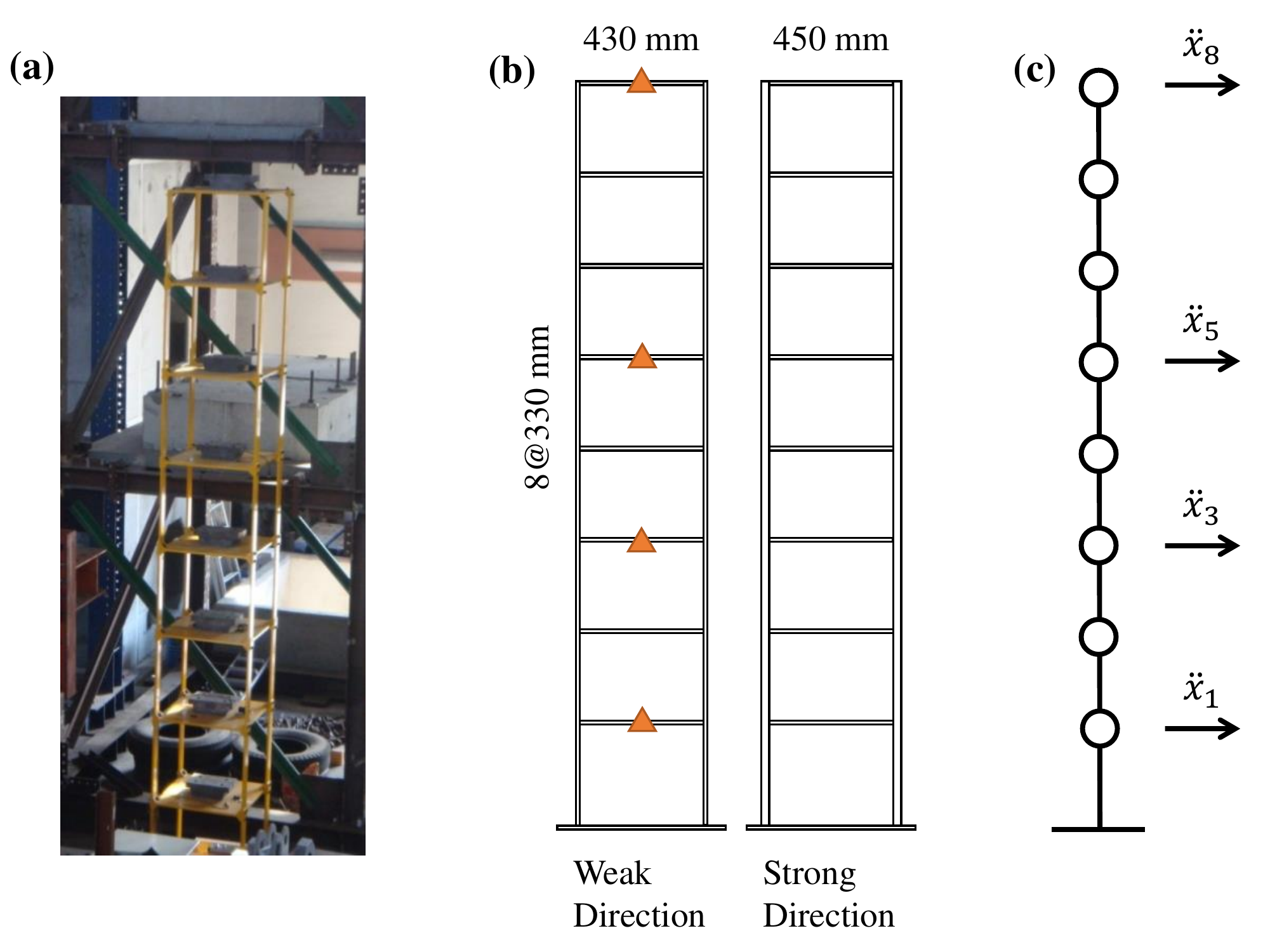}
\caption{The 8-story steel frame: (a) the experimental structure; (b) lateral views of the structure; (c) the condensed shear-type model. Seismic motions are applied along the weak direction. The triangles indicate the accelerometers.}
\label{Fig:SteelFrame}
\end{figure}

The performance of different regularization techniques is also compared with the damage identification result shown in Figure \ref{Fig:Truss_ReguComparison}. Assuming an exact intact model, we apply STLS, LASSO and ridge regression, which represent $\ell_0$, $\ell_1$ and $\ell_2$ regularization, to find the best approximation for the sparse damaged model from the 10\% RMS noise measurement. \zcedit{Ridge regression \cite{Marquardt}, also known as Tikhonov regression \cite{Tikhonov}, is a penalized multicollinearity in data by adopting an $\ell_2$ penalty.} It turns out STLS correctly identifies both the damage location and extents. LASSO comes second very closely, given that there are several false positives\zcedit{, especially for Element 8 which could be distracting}. The Ridge regression provides over-smooth result\zcedit{. It is quite clear that, despite the two principle damages in Elements 1 and 15, we are very likely to conclude from Ridge regression result that Elements 8, 22, 27, 29 and 30 all have notable damages. More importantly, the stiffness variation in Element 27 is incorrectly identified.} \zcedit{These indicate} that $\ell_2$ regularization tends to regularize every element and \zcedit{balances weights across all elements. Consequently, it is not as competitive as STLS or LASSO} for sparse damage identification. \zcedit{Overall, the relative $\ell_2$-norm identification errors for STLS, LASSO and Ridge are 3\%, 26\% and 41\%, respectively.} This comparison demonstrates the numerical advantage of $\ell_0$ regularization for solving sparse regression problems such as damage detection.

\subsection{Experimental Example: An 8 DOF Steel Frame}
A shake table test of an 8-story steel frame model (see Figure \ref{Fig:SteelFrame}) performed in the National Center for Research on Earthquake Engineering in Taiwan \cite{Yu, SunMSSP2016, Chen} is adopted to further verify the proposed model updating and damage identification approach. The total height of the structure is 8 $\times$ 330 mm and the size of diaphragms is 430 mm by 450 mm. The story mass is about 75 kg after counting in a 50 kg steel block on each floor for stabilization. The story lateral stiffness is estimated to be 180 kN/m. Due to its dominant shear component, we approximately condense an ETABS nominal model into a shear-type structure. The condensed stiffness matrix \cite{Sun2017,SunSCHM2018} is obtained by applying a unit lateral load at each floor and then inverting the resulting flexibility matrix. Notwithstanding that the condensed model has discrepancy as compared to the actual system. Moreover, structural damages were intentionally created by loosening bolts connecting adjacent columns. We consider two damage cases: in Case 1, connection bolts on the first floor are loosened; in Case 2, bolts on the first and the second floors are loosened.

Fixed on a hydraulic uniaxial shake table, the structure was monitored under 9 earthquake records in the weak direction (subjected to El Centro, Chi-Chi and Kobe earthquake excitation under different scales). Acceleration time histories were recorded on each floor. Although complete data are available, only the acceleration at the first, third, fifth and eighth floors along with the ground motion are used herein. Once again, we use \zcedit{OKID/ERA} \cite{Juang,Juang1985} to identify the modal frequencies and shapes for the first three modes.

\begin{figure*}[h]
\centering
\includegraphics[width=0.70\textwidth]{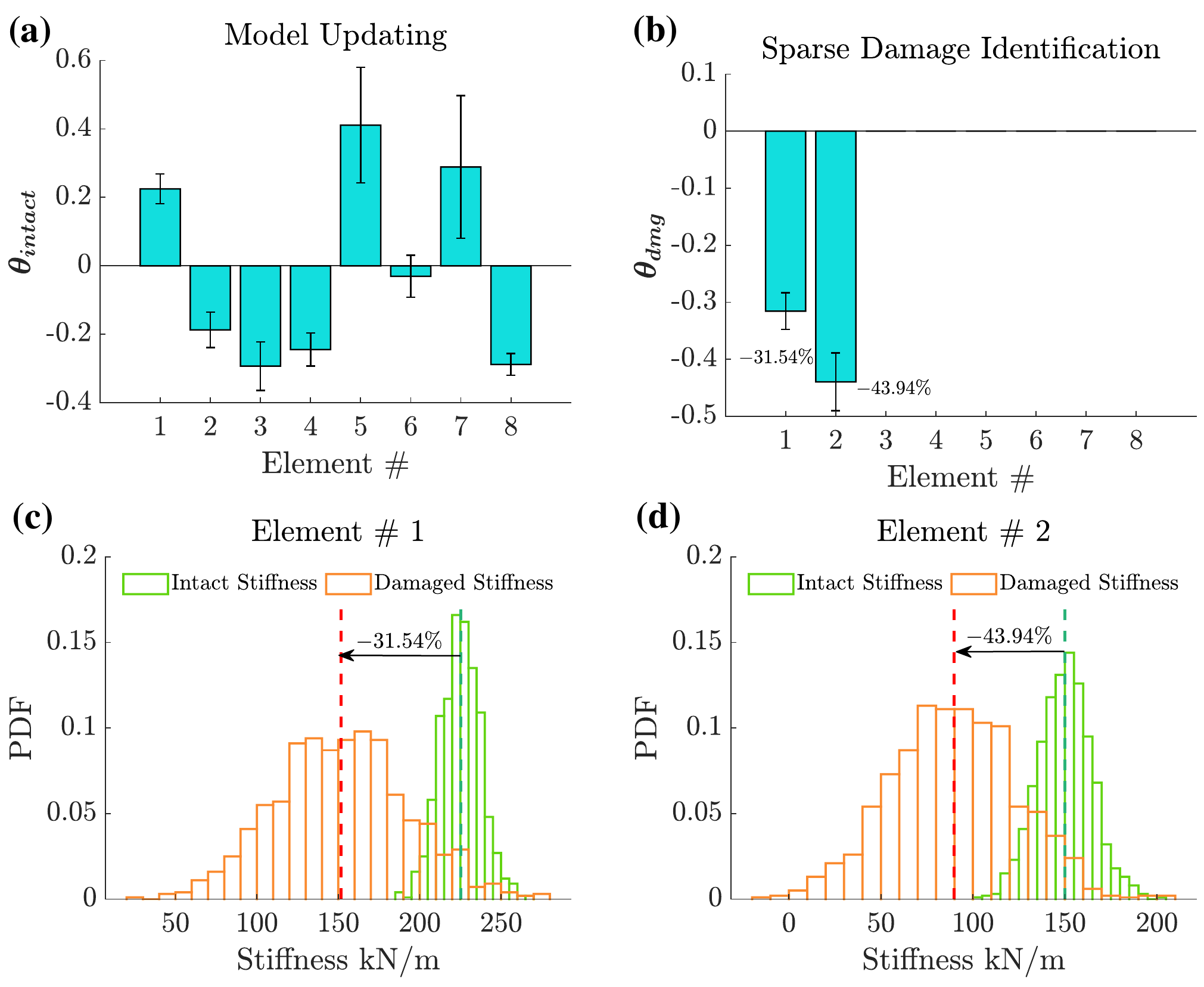}
\caption{Model updating and sparse damage case 1 for the steel frame: (a) stiffness variation $\boldsymbol{\theta}_{intact}$ from the initial model to the intact model; (b) stiffness variation $\boldsymbol{\theta}_{dmg}$ from the intact model to the damaged model in case 1;(c) stiffness distribution shift by 31.54\% in Element 1; (d) stiffness distribution shift by 43.94\% in Element 2.}
\label{Fig:ShakeTable_case1}
\end{figure*}

We first update the stiffness variations parameters for the intact model using $\ell_2$ Bayesian learning. Figure \ref{Fig:ShakeTable_case1}(a) shows the distribution of the updated parameters with 95\% confidence intervals. It can be observed that although Elements 5 and 7 have relatively larger variance, the rest estimates are more reliable. Hinged on the updated mean values of $\boldsymbol{\theta}_{intact}$, we perform damage identification using the proposed STLS regression approach. Figure \ref{Fig:ShakeTable_case1}(b)-(d) shows the identified stiffness reduction of 31.54\% and 43.94\% in Elements 1 and 2 for Case 1. This aligns with our expectation that only the stiffness of the first and the second columns should have major reduction due to the bold loosening. Even though Elements 5 and 7 have large variance in model updating, we still obtain a satisfactory sparse damage identification result. On one hand, the result illustrates that our proposed framework has an agreeable prediction of the mean value in this case; on the other hand, it turns out that our STLS method can identify the essential pattern from the measurements and leave out minor redundancy due to noise and some modeling errors. Figure \ref{Fig:ShakeTable_case1}(c) and (d) show the quantified uncertainties for stiffness parameters with reduction.  

\begin{figure*}[h]
\centering
\includegraphics[width=0.70\textwidth]{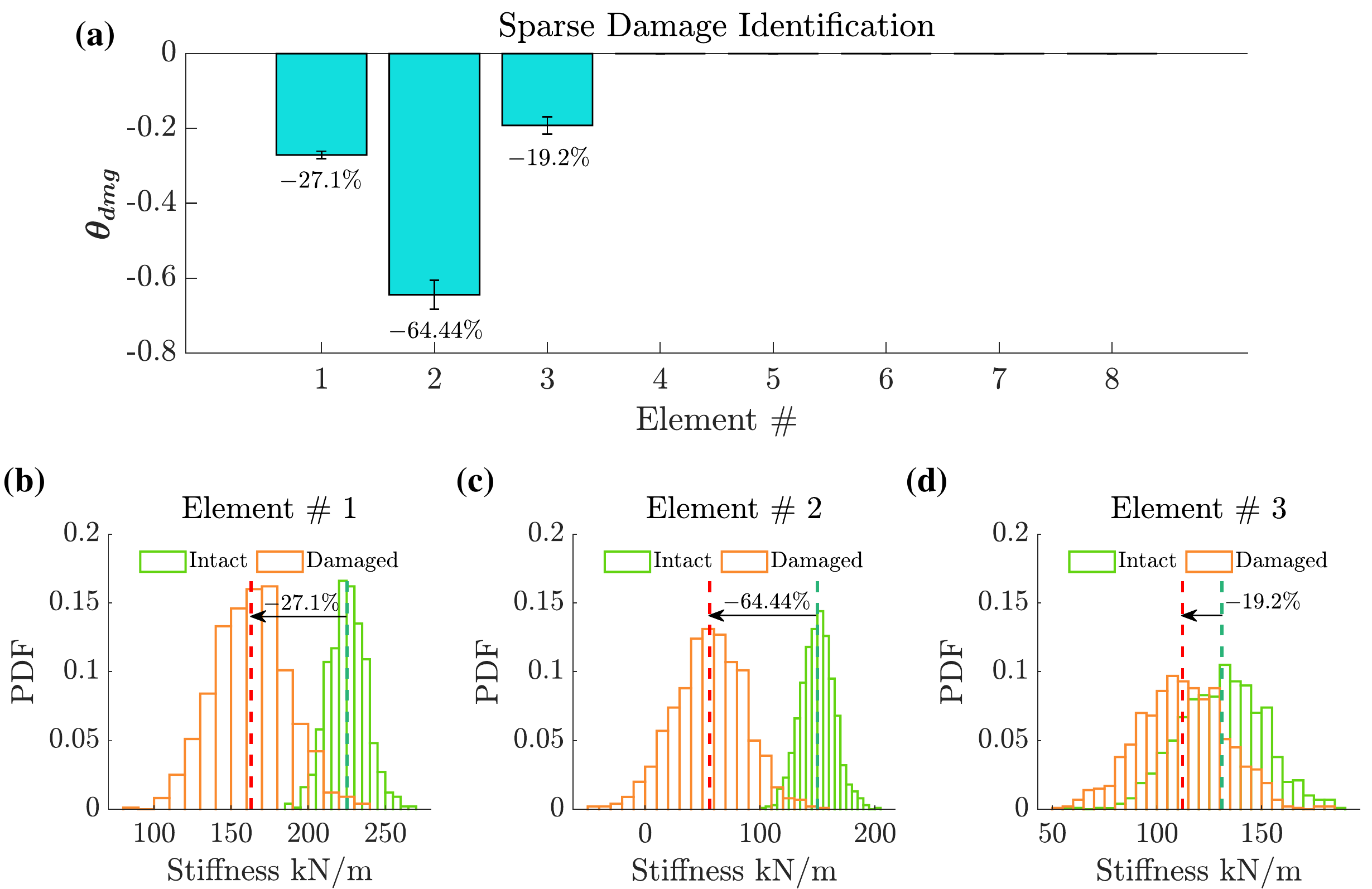}
\caption{Sparse damage case 2 for the steel frame: (a) stiffness variation $\boldsymbol{\theta}_{dmg}$ from the intact model to the damaged model in case 2;(b) stiffness distribution shift by 27.1\% in Element 1; (c) stiffness distribution shift by 64.44\% in Element 2; (d) stiffness distribution shift by 19.2\% in Element 3.}
\label{Fig:ShakeTable_case6}
\end{figure*}

The damage identification results for Case 2 are summarized in Figure \ref{Fig:ShakeTable_case6}, where again their mean values meet our expectation (the loosened bolts would cause stiffness reduction in the first three stories) and the 95\% confidence intervals of the first and third elements are relatively small showing a strong identification confidence. It can be seen from Figure \ref{Fig:ShakeTable_case6} that the stiffness reduction rates for the first three elements are 27.1\%, 64.44\% and 19.2\%. The second element suffers the most because both of its ends were loosened. The probabilistic distributions of identified stiffness parameters are given in Figure \ref{Fig:ShakeTable_case6}(b)-(d). 

\section{Conclusion} \label{Conclusion}
This paper develops a novel two-stage sensitivity analysis-based computational framework for both model updating and sparse damage identification. In particular, an $\ell_2$ Bayesian learning method is developed for intact model updating and uncertainty quantification. The updated model then serves as a baseline for damage identification. A sparse representation pipeline built on a quasi-$\ell_0$ method (STLS regression) is presented for \zcedit{sparse} damage localization and quantification. \zcedit{While the smooth nature of $\ell_2$ Bayesian learning makes it preferable for largely populated damages, there are many cases (deterioration of connection rigidity) where sparse damages only occur at distinct locations, justifying the necessity for sparse identification.} Nevertheless, a critical issue of STLS lies in how to choose the thresholding parameter which is very problem-dependent. An inappropriate selection of such a parameter likely leads to biased identification. To address this fundamental issue, Bayesian optimization together with cross validation is developed to intelligently fit STLS with data, which saves the computational cost of hyperparameter tuning and produces more reliable identification result. The proposed framework is verified by three examples (both numerical and experimental), including a 10-story shear-type building, a complex truss structure, and a shake table test of an eight-story steel frame. In all cases, the $\ell_0$ Bayesian learning method can reliably estimate the probable stiffness variation, while the STLS regression can localize the sparsely distributed damaged members and quantify the damage extents with high accuracy. The encouraging results set forward our future work to be focused on real-world applications of the proposed methodology.

\section{Acknowledgment}
We would like to thank the National Center for Research on Earthquake Engineering in Taiwan for sharing the shake-table test data.

\section{Declaration of Conflicting Interests}
The Authors declare that there is no conflict of interest.

\end{document}